\documentclass[floatfix,showpacs,amsmath,amssymb,letterpaper,groupaddresses,superscriptaddress]{article}
\usepackage{graphicx}
\usepackage{latexsym}
\usepackage{amsmath,amssymb}

\usepackage{epsfig}
\usepackage{graphicx}

\usepackage{graphicx}
\usepackage{amssymb,amsmath,times}

\usepackage{color}

\def\a{\alpha}
\def\b{\beta}

\def\m{\mu}
\def\n{\nu}
\def\be{\begin{equation}}
\def\ee{\end{equation}}
\def\ba{\begin{eqnarray}}
\def\ea{\end{eqnarray}}
\def\la{\langle}
\def\ra{\rangle}
\def\a{\alpha}

\def\b{\beta}

\def\m{\mu}
\def\n{\nu}

\def\h{\hskip 1cm}
\def\hh{\hskip 2cm}
\def\lo{\longrightarrow}

\usepackage{epsfig}
\usepackage{graphicx}

\begin{document}
\begin{center}
{\Large \bf Characterization of qutrit channels, in terms of their covariance and symmetry properties}\\
\vspace{1cm}
Vahid Karimipour\footnote{email: vahid@sharif.edu, corresponding
author}\h Azam Mani\footnote{email:mani-azam@physics.sharif.edu}\h
Laleh Memarzadeh \footnote{email: memarzadeh@sharif.edu}\\
\vspace{5mm}

\vspace{1cm} Department of Physics, Sharif University of Technology,\\
P.O. Box 11155-9161,\\ Tehran, Iran
\end{center}
\vskip 3cm

\begin{abstract}
We characterize the completely positive trace-preserving maps on
qutrits, (qutrit channels) according to their covariance and
symmetry properties. Both discrete and continuous groups are
considered. It is shown how each symmetry group, restricts
arbitrariness in the parameters of the channel to a very small
set. Although the explicit examples are related to qutrit
channels, the formalism is sufficiently general to be applied to
qudit channels.
\end{abstract}
\vskip 1cm Keywords: Quantum channels, covariant channels, symmetry.\\
PACS Numbers: $03.67$.-a, $03.65$.Aa, $03.67$.Hk
% quantum information , Quantum systems with finite Hilbert space, Quantum communication

\section{Introduction}\label{intro}
The problem of characterization of quantum channels is one of the
interesting and important problems of quantum information
science. On the physical side, a quantum channel is the most
general physical process that a quantum system can undergo
\cite{sudar, kraus}, when initially it is only classically
correlated \cite{ce} with its environment. (Non-positivity of the
corresponding map indicates the presence of an initial quantum
correlation between the system and the environment \cite{ce}). On
the mathematical side, this problem is equivalent to
characterizing completely positive maps. One of the basic results
is that, any channel has an operator sum representation, that is,
for any such channel ${\cal E}$ can be represented as
\begin{equation}\label{}
  {\cal E}(\rho)=\sum_k A_k \rho A_k^{\dagger},
\end{equation}
where the condition $\sum_k A_k^{\dagger}A_k=I$ guarantees that
the channel is trace preserving. What renders the
characterization of such channels difficult, is due to the
non-uniqueness of the kraus representation. That is any channel
can have infinitely many equivalent Kraus representation. Any
two such representations, one by the Kraus set $\{A_a\}$, and the
other by $\{B_a\}$ are related by a unitary transformation \cite{niel}, that
is, there is a unitary matrix $\Omega$ such that
\begin{equation}\label{unit}
  B_a=\sum_b\Omega_{ab}A_b.
\end{equation}
Despite this difficulty, a complete characterization of qubit
channels is at hand thanks to the works of \cite{Ruskai} and \cite{Fujiwara}.
We now know that any qubit channel is of the following form
\begin{equation}\label{qubitform}
  {\cal E}(\rho) = U{\cal E}_{{\bf t},\Lambda}(V\rho V^{-1})
  U^{-1},
\end{equation}
where ${\cal E}_{{\bf t},\Lambda}(\rho)$ maps the vectors in the
Bloch sphere as follows ${\bf r}\lo {\bf r'}={\bf {\Lambda}}{\bf
r}+{\bf t}$, where ${\bf t}$ is a three-dimensional vector,
$\Lambda = diag (\lambda_1, \lambda_2, \lambda_3)$ and the vector
$\overrightarrow{\lambda}  = (\lambda_1, \lambda_2, \lambda_3)$ is
contained in a tetrahedron spanned by the four corners of the
unit cube with $\lambda_1\lambda_2\lambda_3=1$ \cite{Fujiwara, King}. For higher
dimensional channels no such characterization has yet been made,
even for the simplest case of qutrit channels. The difficulty with
higher dimensional channels, lies with both the large number of
parameters needed for description of states and with the absence
of a proper geometrical representation for states, as Bloch sphere for qubits.
 In fact the positivity of the density matrix for qudits
 does not lead to simple conditions and simple geometry
 \cite{kimura, khaneja, mendas, boya,kryszewski}.\\

To see the root of the problem for characterization of higher
dimensional channels, let us take the simplest example of $d=3$.
A qutrit state can be described by a matrix $\rho=\frac{1}{3}(I+
{\sqrt3} {\bf r}\cdot {\bf \Gamma})$, where $\Gamma_{1}$ to
$\Gamma_8$ are the Gell-Mann matrices (a Hermitian basis for the
Lie algebra of $su(3)$), and ${\bf r}$ is an 8-dimensional vector
with ${\bf r}\cdot{\bf r}\leq 1$. Therefore the role of the
two-dimensional Bloch sphere is played by a 7 dimensional sphere,
however the point and the root of difficulty is that not all the
vectors inside this sphere represent physical states, i.e.
positive states $\rho$. In general any completely positive map
induces an affine transformation on the vector ${\bf r}$ in the
form ${\bf r}\lo {\bf r'}={\bf \Lambda}{\bf r}+{\bf t}$. For
qubits, local unitary changes of basis for the input and output
spaces, diagonalize the matrix $\Lambda$, hence lead to the form
(\ref{qubitform}), however for qutrits, the parameters of these
local unitaries ($SU(3)\otimes SU(3)$) are much less than those
of ${\bf \Lambda}$ (which is equal to $8^2$) and this
diagonalization is not possible. This renders a complete
characterization of qutrit channels difficult.\\

Nevertheless, the problem of characterizing qutrit maps, has been
tackled from different points of view. In \cite{checinska}, a
special class of qutrit channels whose matrix $\Lambda$ of affine
transformation is already diagonal has been studied to find the
complete positivity constraints on qutrit channels. In
\cite{Byrd}, general properties of the affine map on polarization
vectors, obtained from completely positive maps on d-level states
(qudits) ,has been obtained. In particular special classes of
qudit channels with special class of Kraus operators, i.e.
unitary, Hermitian, and orthogonal, have
been investigated. \\

In this work, we follow a different approach to categorize qutrit channels
in terms of their covariance or symmetry property under some discrete or continuous
symmetry group. Our study like the ones in \cite{Ruskai} and \cite{Fujiwara}
does not provide an exhaustive characterization of qutrit
channels, but provides a framework for a better understanding of
the space of qutrit channels. This study sheds light on some
aspects of these channels which we hope along with other studies
will pave the way for a full characterization of these channels
in the future.\\

The structure of this paper is as follows: In section (\ref{Prelim}), we
provide the preliminary notations about completely positive
channels and set up our notations and conventions. In sections
(\ref{covarance}) and (\ref{symmetry}), we explain the concept of
 covariance and symmetry of quantum channels and
derive the conditions that these properties impose on the Kraus
operators of such channels. In sections (\ref{discexam}) and
 (\ref{contexam}) we study in detail several classes of examples,
  both for discrete and for continuous groups.
Although we mainly study the qutrit channels, the formalism that
we explain and indeed some of the examples we suggest are
general and apply for quantum channels on d-dimensional states or
qudits.\\

\section{Preliminaries}\label{Prelim}

The vector space of $d$ dimensional complex square matrices is
denoted by $M_d$. With the definition of the Hilbert-Schmidt inner
product $\la A, B\ra = tr(A^{\dagger}B)$ it becomes a Hilbert
space of dimension $d^2$. Let ${\ \Gamma_\m\  (\m=0,\cdots, d^2-1})$
be an orthonormal hermitian basis for this vector space
\begin{equation}\label{bas}
\la \Gamma_\mu,
\Gamma_\nu\ra=tr(\Gamma_\mu^{\dagger}\Gamma_\nu)=\delta_{\mu,\nu}.
\end{equation}
Any unit-trace matrix $\rho\in M_d$ can be expanded as

\begin{equation}\label{rho}
 \rho=\sum_{\mu=0}^{d^2-1} \rho_\mu \Gamma_\mu = \frac{1}{d}I +
 \sum_{i=1}^{d^2-1} \rho_i \Gamma_i,
\end{equation}
where

\begin{equation}\label{rho-mu}
  \rho_\m = \la \Gamma_\mu, \rho\ra = tr(\Gamma_\m^{\dagger}\rho),
\end{equation}
and for the first component we have
\begin{equation}\label{bas0}
\Gamma_0=\frac{1}{\sqrt{d}}I, \h \rho_0 = \frac{1}{\sqrt{d}}.
\end{equation}
This, together with (\ref{bas}) implies that the matrices
$\Gamma_{i}$ are traceless. Note that we use the Latin indices
$i,j,\cdots $ for $\mu\ne 0$. For a density matrix, the condition
$tr(\rho^2)\leq 1$, constrains the polarization vector to lie
within a sphere of radius $\sqrt{1-\frac{1}{d}}$. However not all
the vectors in this sphere represent density matrices, due to the
extra condition of
the positivity of the density matrix \cite{khaneja}. \\
\\
A completely positive map ${\cal E}:M_d\lo M_d$, i.e. a channel
is represented as
\begin{equation}\label{chn}
\rho'={\cal E}(\rho)=\sum_a A_a\rho A_a^{\dagger},
\end{equation}
where the set of operators $\{{\cal A}_a\}$ are called Kraus
operators \cite{kraus}. It is well-known that any two Kraus representations of
a channel are connected by a unitary transformation \cite{niel}, that is if
$\{A_a\}$ and $\{B_a\}$ are Kraus operators of the same channel
${\cal E}$, then there exists a unitary matrix $\Omega$ such that
\begin{equation}\label{two}
  A_a=\sum_b \Omega_{ab} B_b.
\end{equation}
In writing this transformation, one assumes that the two sets of
Kraus operators are of equal size, since one can always add zero
Kraus operators without changing the channel.\\
\\
To find how the polarization vector transforms under such a map,
we note that

\begin{equation}\label{Erhocomp}
{\cal E}(\rho) = \sum_\mu \rho'_\mu \Gamma_\m = \sum_{a,\n}
\rho_\n A_a \Gamma_\n A_a^{\dagger},
\end{equation}
from which we obtain
\begin{equation}\label{Lambda}
  \rho'_\m = \Lambda_{\m,\n}\rho_\n,
\end{equation}
in which
\begin{equation}\label{LambdaComp}
  \Lambda_{\m,\n}=\la \Gamma_\m, {\cal E}(\Gamma_\n)\ra=\sum_a tr (\Gamma_\m^{\dagger}A_a \Gamma_\n A_a^{\dagger}).
\end{equation}
For a trace-preserving map ($tr({\cal E}(\rho)=tr(\rho)$), we
have $\sum_a A_a^{\dagger}A_a=I$ and for a unital ($ {\cal
E}(I)=I$), we have $\sum_a A_a A_a^{\dagger}=I$. Therefore using
(\ref{LambdaComp}), we find
\begin{eqnarray}\label{lambda0i}
  &&\Lambda_{0i}=0 \h {\rm if \ \ {\cal E}\ \  is\  trace-preserving\ },\cr
  &&\Lambda_{i0}=0 \h {\rm if \ \ {\cal E} \ \ is\  unital\ }.
\end{eqnarray}
In both cases, we always have $\Lambda_{00}=1$. For
trace-preserving maps the form of the matrix $\Lambda$ becomes
\begin{equation}\label{affine}
  \left(\begin{array}{c}\frac{1}{\sqrt{d}}\\ {\bf \rho'}\end{array}\right)=
  \left(\begin{array}{cc} 1 & 0 \\ {\bf T} & {\bf \Lambda}\end{array}\right)\left(\begin{array}{c}\frac{1}{\sqrt{d}}\\ {\bf
  \rho}\end{array}\right),
\end{equation}
which means the action of the channel on the vector ${\bf
\rho}:=(\rho_1, \cdots \rho_{d^2-1})$ is an affine map, i.e.
${\rho}\lo {\bf \Lambda} {\bf \rho}+{\bf T}$. An analysis of the
completely positive maps based on these affine maps, with
examples of qutrit channels has been carried out in \cite{Byrd}.\\

As pointed out in the introduction, we want to study qutrit channels
from the perspective of their covariance and symmetry properties.
To this end, we begin the next section with general remarks on
covariance of completely positive maps.

\section{Covariant Channels} \label{covarance}
Let $G$ be a group with $D^{^{(1)}}$ and $D^{^{(2)}}$ being its
representations on $M_d$. We call the channel ${\cal
E}$ covariant under group $G$ with respect to these two
representations if for all $g\in G$, the following relation holds
\begin{equation}\label{covdef}
  D^{^{(2)}}(g){\cal E}(\rho)D^{^{(2)}}(g^{-1})={\cal E}(D^{^{(1)}}(g)\rho
  D^{^{(1)}}(g^{-1})).
\end{equation}
This means that a transformation of the input density matrix by
$D^{^{(1)}}(g)$ results in a corresponding transformation of the
output density matrix by $D^{^{(2)}}(g)$. Note that although the
input and output density matrices may be of the same dimensions,
they may transform according to different representations of the
symmetry group. We will see examples in section (\ref{Su3sub}),
when we discuss channels covariant under the $Su(3)$ group which has two inequivalent 3-dimensional representations.\\

Besides this appealing property, such channels offer a lot of
convenience, when we want to calculate their one-shot classical
capacities $C^1({\cal E})$ \cite{holevo}. To see this, we note
\begin{equation}\label{C1}
  C^1({\cal E})=Sup_{\{p_i,\rho_i\}}\chi({\cal E},\{p_i,\rho_i\})=
  Sup_{\{p_i,\rho_i\}}\left[S(\sum_i p_i {\cal E}(\rho_i))-\sum_i p_i
  S({\cal E}(\rho_i))\right].
\end{equation}
Here, $\chi$ and $S$ denote respectively the Holevo quantity and
the von Newmann entropy, and the maximization is performed over
all input ensembles $\{p_i,\rho_i\}$. For a covariant channel,
with irreducible representation $D^{^{(2)}}$, the problem of finding
 an optimal ensemble reduces to the problem of finding the minimum
output entropy state $\rho^*:=|\psi^*\ra\la \psi^*|$, i.e. the
pure state which minimizes the second term $S({\cal E}(\rho^*))$ \cite{holevo-cov,macchia}.
(Purity of this state follows from the convexity of the output
entropy) Once this state is found, we can maximize the first term
and hence the Holevo quantity itself by taking as the input
ensemble the uniformly distributed ensemble of $\rho(g) =
D^{^{(1)}}(g)\rho^*D^{^{(1)}}(g^{-1})$. In case that $D^{^{(2)}}$
is an irreducible representation, one finds that \cite{groupthe}
\begin{equation}\label{sumg}
  \frac{1}{|G|}\sum_{g\in G}D^{^{(2)}}(g){\cal E}(\rho^*)D^{^{(2)}}(g^{-1}) =\frac{1}{d} I,
\end{equation}
which means that the first term will take its maximum value for
such an ensemble. Therefore the one-shot classical capacity of
covariant channels will be found to be
\begin{equation}\label{one-shot}
  C({\cal E})=\log_2 d - S({\cal E}(\rho^*)).
\end{equation}
Here we see how covariant property of the channel reduces a
problem which involved an optimization over a large parameter
space (in this case the space of input ensembles) to the much
simpler problem of finding the minimum output entropy of the
channel. The input space that we have to search for can be
further reduced if the channel has some kind of symmetry, like
$${\cal E}(\rho)={\cal E}(h\rho h^{-1}),
$$ where $h$ is an element of a group $H$ . In such a case
it is enough to search over a subset of states which are invariant
under $H$. Furthermore, due to the convexity of the output entropy,
the search can be restricted to be over pure states with that symmetry.\\

Before going to a systematic discussion for constructing
covariant maps, let us note the simplest examples. Clearly the
identity channel ${\cal E}_I(\rho)=\rho$, is a CPT map which is
covariant under any group of transformations, with
$D^{(1)}=D^{(2)}.$ The completely mixing CPT map ${\cal
E}_0(\rho)=\frac{1}{d}tr(\rho)I$ is covariant under any
transformation group with arbitrary representations $D^{(1)}$ and
$D^{(2)}$. As a less trivial example, consider the map ${\cal
E}_T(\rho)=\frac{1}{d+1}(tr(\rho)I + \rho^T)$, where $\rho^T$
denotes the transpose of $\rho$. Although transposition by itself
is not a completely positive map, the above convex combination
with the mixing map is a CPT, since it has a Kraus representation
in the form:

\begin{equation}\label{ET}
  {\cal E}_T(\rho)=\frac{1}{2(d+1)}\sum_{i,j}A_{ij}\rho
  A_{i,j}^{\dagger},
\end{equation}
where $A_{ij}=|i\ra\la j|+|j\ra\la i|.$ This channel is covariant
under any group where $D^{(1)}(g)=g$ and $D^{(2)}(g)=g^*$. The
convex combination of any two channels which are covariant under
the same representations of a given group, is also covariant
under the same group with the same representations. For example
consider a qudit channel. Considering the qudit as the state of a
spin-$(d-1)/2$ particle, the group $SO(d)$ has a natural action
on it in the form of rotation $|\psi\ra\lo R|\psi\ra$, or
$\rho\lo R\rho R^T$. Noting that the representation $R$ is real,
$R=R^*$, we find that the following qudit channel is covariant
under the rotation group $SO(d)$:

\begin{equation}\label{R}
  {\cal E}(\rho)=a {\cal E}_I (\rho) + b {\cal E}_0 (\rho) + (1-a-b) {\cal
  E}_T(\rho),
\end{equation}
or by redefining the parameters,
\begin{equation}\label{R}
  {\cal E}(\rho)=\alpha \ tr(\rho)\ I  + \beta \ \rho  + \gamma \ \rho^T,
\end{equation}
with  $\alpha d + \beta + \gamma =1.$\\
\\
In order to characterize covariant channels in a systematic
way, it is best to consider the Kraus representation of such
channels in terms of which, equation (\ref{covdef}) reads
\begin{equation}\label{covkraus}
  D^{^{(2)}}(g)\sum_a A_a \rho A_a^{\dagger}D^{^{(2)}}(g^{-1})=\sum_a A_a
  (D^{^{(1)}}(g)\rho D^{^{(1)}}(g^{-1}))A_a^{\dagger},
\end{equation}
or equivalently
\begin{equation}\label{covkrauscombine}
  \sum_a A_a\rho A_a^{\dagger}=\sum_a D^{^{(2)}}(g^{-1})A_a
  (D^{^{(1)}}(g)\rho D^{^{(1)}}(g^{-1}))A_a^{\dagger}D^{^{(2)}}(g).
\end{equation}
However according to (\ref{two}) any two different Kraus
representations of a channels are necessarily related by
a unitary transformation, which implies that
\begin{equation}\label{covfinal}
   D^{^{(2)}}(g^{-1})A_a D^{^{(1)}}(g)=\sum_b \Omega_{ab}(g) A_b.
\end{equation}
Repeating this relation for two different group elements $g$ and
$g'$ and combining the two we find that $$\sum_b
\Omega_{ab}(g)\Omega_{bc}(g')=\Omega_{ac}(gg'),$$ which means that
$\Omega$ is a unitary representation of the group $G$. The
dimension of this representation is the same as the number of
Kraus operators. Moreover from (\ref{covfinal}) one finds that

\begin{equation}\label{covfinaldagger}
   D^{^{(1)}}(g^{-1})A_a^{\dagger}
  D^{^{(2)}}(g)=\sum_b \Omega^*_{ab}(g) A_b^{\dagger}.
\end{equation}
Combining this relation with (\ref{covfinal}), we find that

\begin{eqnarray}\label{commutation}
   && [D^{^{(1)}}(g), \sum_a A_a^{\dagger}A_a]=0, \cr
   && [D^{^{(2)}}(g), \sum_a A_aA_a^{\dagger}]=0.
\end{eqnarray}\\

The first relation implies that if $D^{^{(1)}}$ is an irreducible
representation, then according to the Schur's Lemma \cite{groupthe}, $\sum_a
A_a^{\dagger}A_a \propto I $ and hence by appropriate
normalization, the map ${\cal E}$ can be made
trace-preserving. Similarly the second relation implies that if
$D^{^{(2)}}$ is an irreducible representation, $\sum_a
A_aA_a^{\dagger}$ $ \propto I $ and hence by appropriate
normalization, the channel
${\cal E}$ can be made unital.\\

To find channels with covariant property we should find a set of
Kruas operators satisfying equation (\ref{covfinal}), where
$D^{^{(1)}}$, $D^{^{(2)}}$ and $\Omega$ are the representations
of the group. There are many different choices for the
representations of a group, but they may
give the same or equivalent maps. By the following remarks we
introduce the strategy
by which we can restrict our attention to limited number of representations for $D^{^{(2)}}$, $D^{^{(1)}}$ and $\Omega$.\\

\textbf{Remark 1:} For $\Omega$ it is enough to consider all the
irreducible representations of the group. For each irreducible
representation $\Omega^{(i)}$ we can find a set of Kraus operators
satisfying
\begin{equation}\label{remark1}
  D^{^{(2)}}(g^{-1})A^{(i)}_a D^{^{(1)}}(g) = {\Omega}^{(i)}_{ab}(g)A^{(i)}_b, \h
\end{equation}
where ${\Omega}^{(i)}$ is an irreducible representation of
$G$  ($i=1\cdots K$, labels the $K$ different irreducible
representation of $G$). Therefore we can define $K$ sets of Kraus
operators or equivalently $K$ channels which are covariant under
$G$
\begin{equation}\label{Ei}
  {\cal E}^{(i)} (\rho) = \sum_{a} A^{(i)}_a \rho A^{(i)^{\dagger}}_a.
\end{equation}
It is clear that any convex combination of these maps is also covariant under the action of
the group $G$. Therefore the overall solution can be represented by
\begin{equation}
{\cal E}=\sum_{i=1}^K\lambda_i{\cal E}^{(i)},
\end{equation}
and it is enough to consider irreducible representations of
$\Omega$ without loss of  generality. Note that the
representations which we choose for $D^{^{(1)}}$ and $D^{^{(2)}}$
need not be irreducible. In fact we will see explicit cases of
reducible representations in the examples which follow the
general formalism. \\

\textbf{Remark 2}: Let the representations $D^{^{(1)}}$ and
$D^{^{(2)}}$ be respectively equivalent to the representations
$D'_1$ and $D'_2$, i.e. let $U$ and $V$ be unitary operators acting on
$M_d$ such that for all $g\in G$

$$D'_1 (g) = UD^{^{(1)}}(g) U^{\dagger}, \h D'_2 (g) = VD^{^{(2)}}(g) V^{\dagger},$$
If the channel ${\cal E}$ is covariant with respect to
$D^{^{(1)}}$ and $D^{^{(2)}}$, the channel ${\cal E'}:=V\circ
{\cal E}\circ U$ will be covariant with respect to $D'_1$, and
$D'_2$. The channel ${\cal E}'$ is defined as
\begin{equation}\label{E'}
  {\cal E}'(\rho)=V{\cal E}(U\rho U^{-1})V^{-1}.
\end{equation}
Therefore without loss of generality we consider only
non-equivalent representations for $D^{^{(1)}}$ and $D^{^{(2)}}$.

%\textbf{Remark 3:} We now ask what happens if the representation $D$ is reducible,
%i.e. if for all $g\in G$ we have
%\begin{equation}\label{D(g)red}
%  D(g) = \left(\begin{array}{cc} E(g) & 0 \\ 0 & F(g)\end{array}\right)
%\end{equation}
%
%\begin{equation}\label{sumkrausred}
%  \sum_a A_a^{\dagger}A_a = \left(\begin{array}{cc} A & B \\ C & D \end{array}\right)
%\end{equation}
%
%\begin{equation}\label{ADEF}
%  [A,E(g)]=0, \h [D,F(g)]=I, \ \ \ \forall g\in G
%\end{equation}
%from which we obtain that $A=c_1 I$ and $D=c_2 I$ and
%\begin{equation}\label{BCEF}
%  BF(g)=E(g)B, \h CE(g) = F(g)C, \ \ \ \forall g\in G
%\end{equation}
%
%If $E$ and $F$ are inequivalent representations, then by Shur's
%lemma we obtain that $B=C=0$. Therefore if $D$ is the sum of two
%inequivalent representations, then
%\begin{equation}\label{suminequiv}
%  \sum_a A_a^{\dagger}A_a = \left(\begin{array}{cc} c_1 I  & 0 \\ 0 & c_2 I
%  \end{array}\right), \ \ \ \ \ \sum_a A_a A_a^{\dagger} = \left(\begin{array}{cc} D^{^{(1)}} I  & 0 \\ 0 & D^{^{(2)}} I
%  \end{array}\right)
%\end{equation}

Before embarking into an investigation of such maps for qutrit
channels, we first study in general another important property,
namely symmetry of a channel under a group of transformation.
\section{Symmetric Maps}\label{symmetry}
A symmetric channel has the property that for elements $g\in G$
of a group the following property holds

\begin{equation}\label{symdef}
  {\cal E}(D(g)\rho D(g^{-1})) = {\cal E}(\rho).
\end{equation}

In this case, we say that ${\cal E}$ is symmetric with respect to the
representation $D$ of the group. An example of such a channel is
the bit-flip channel, ${\cal E}(\rho)= (1-p)\rho + p \ X \rho X$, with $p=\frac{1}{2}$ which is symmetry under
the group $G=\{I, X\},$
where $X$ is the bit-flip operator.\\
\\
In terms of the Kraus representations, this is equivalent to
\begin{equation}\label{symkrausdef}
  \sum_a A_a D(g)\rho D(g^{-1}) A_a^{\dagger} = \sum_a A_a \rho
  A_a^{\dagger},
\end{equation}
or according to (\ref{two})
\begin{equation}\label{symkrauscond}
  A_a D(g) = \sum_b \Omega_{ab}(g) A_b,
\end{equation}
where again $\Omega$ is a representation of $G$. Obviously
if the channel ${\cal E}$ is symmetric with respect to the
representation $D$, then the channel ${\cal E}\circ {\cal U}$
will be symmetric with respect to the equivalent representation
$D'=UDU^{-1}$ . Therefore we only need to consider the completely
positive maps which are symmetric under inequivalent
representations of a given group.\\

In the next sections we investigate in more detail examples of
completely positive maps which are covariant or symmetric
under various discrete or continuous groups, with an emphasis on
qutrit channels. We split the examples into two separate parts
and consider first the discrete groups and then the continuous
groups.

\section{Examples of covariant qutrit channels under discrete groups}\label{discexam}

In this section we consider several discrete groups and
find classes of channels which are covariant and/or symmetric with respect
to different representations of these groups. First we consider a
cyclic group and then proceed to other Abelian and Non-Abelian
discrete groups.

\subsection{The Abelian case: Cyclic groups}
Consider a cyclic group of order $n$ which is generated by a
single operator called $X$, where $X^n=I$. The order of the group
is generally has nothing to do with the dimension of the Hilbert
space, for example $X$ can be the operator which flips the basis
states $|1\ra$ and $|2\ra$ without affecting the basis state
$|0\ra$ in which case $n=2$, or it can shift all the basis states
by one unit, i.e. $X|i\ra=|i+1, \  {\rm mod \ d}\ra$ in which
case $n=d$. Being Abelian, we know that all the irreducible
representations of this group are one dimensional
\cite{groupthe}. Each such representation is labeled by one
integer $k\in \{0,1,\cdots n-1\}$. According to (\ref{remark1}),
for $\Omega$ we need only take one such representation where $X$
is represented by $e^{\frac{2\pi i k }{n}}$. We have to solve the
following equation for the single Kraus operator $A^{(k)}$

\begin{equation}\label{symmg}
D^{^{(2)}}(X^{-1}) A^{(k)} D^{^{(1)}}(X) = e^{\frac{2\pi i k
}{n}}A^{(k)}.
\end{equation}
To solve this equation we use the eigenvectors of the operators
$D^{^{(1)}}(X)$ and $D^{^{(2)}}(X)$. Let
\begin{equation}\label{Xxi}
  D^{^{(1)}}(X)|\xi_r\ra = e^{\frac{2\pi i r
}{n}} |\xi_r\ra
\end{equation}
and

\begin{equation}\label{Xxi}
  D^{^{(2)}}(X)|\eta_s\ra = e^{\frac{2\pi i s
}{n}}|\eta_s\ra,
\end{equation}
then the general solution of (\ref{symmg}) will be given by
\begin{equation}\label{kraussymmg}
A^{(k)}:=\sum_{l}a_{l,l+k}|\eta_l\ra\la \xi_{l+k}|.
\end{equation}\\

A map whose Kraus operators are of the above form, will be
covariant with respect to the given cyclic group. To make such a
map trace-preserving, the condition $\sum_k
{A^{(k)}}^{\dagger}A^{(k)}$ is imposed which in view of the
explicit form (\ref{kraussymmg}), leads to the condition

\begin{equation}\label{cond1}
  \sum_{n}|a_{n,m}|^2 = 1\ \ \  \forall \ m.
\end{equation}
On the other hand, when the following condition holds, the
channel will be unital

\begin{equation}\label{cond1}
  \sum_{m}|a_{n,m}|^2 = 1\ \ \  \forall \ n.
\end{equation}

\textbf{Example 1:} As a concrete example for qutrits, we consider
the order-3 cyclic group generated by the operator $Z=diagonal(1,
\omega, \omega^2)$ where $\omega^3=1$, and take the
representations $D^{(1)}(Z)=D^{(2)}(Z)=Z$. This is a case where
the order of the cyclic group coincides with the dimension of the
Hilbert space which is 3. In the sequel we consider an example
where these two numbers are not equal. Here we have three Kraus
operators which according to (\ref{kraussymmg}) are given by the
following, where $|0\ra, |1\ra$ and $|2\ra$ are the computational
basis vectors (eigenvectors of $Z$);

\begin{eqnarray}\label{krausZ}
A^{(0)}=a_{00}|0\ra\la 0|+a_{11}|1\ra\la 1|+a_{22}|2\ra\la 2|,\cr
A^{(1)}=a_{01}|0\ra\la 1|+a_{12}|1\ra\la 2|+a_{20}|2\ra\la 0|,\cr
A^{(2)}=a_{02}|0\ra\la 2|+a_{10}|1\ra\la 0|+a_{21}|2\ra\la 1|,
\end{eqnarray}
or in matrix form
\begin{eqnarray}\label{kraus3}
  &&A^{(0)} = \left(\begin{array}{ccc} a_{00} & 0 & 0 \\ 0 & a_{11} & 0 \\ 0 & 0 &
  a_{22}\end{array}\right), \cr
&&A^{(1)} = \left(\begin{array}{ccc} 0 & a_{01} & 0 \\ 0 & 0 &
a_{12}
\\ a_{20} & 0 &
  0\end{array}\right), \cr
&&A^{(2)} = \left(\begin{array}{ccc} 0 & 0 & a_{02} \\ a_{10} & 0
& 0
\\ 0 & a_{21} &
  0\end{array}\right).
\end{eqnarray}
The map will be trace-preserving, if the vectors ${\bf
a}_m=(a_{0m},a_{1m},a_{2m})$ are normalized, and will be a unital
channel if the vectors $\tilde{{\bf a}}_n=(a_{n0},a_{n1},a_{n2})$ are
normalized.\\

\textbf{Example 2:} We now use another type of action, namely the
Hadamard operator
$$H=\frac{1}{\sqrt{d}}\sum_{i,j=0}^{d-1}\omega^{ij}|i\ra\la j|,$$
in which $\omega^{d}=1$. The group which is generated by the Hadamard operators has only
four elements, namely $\{I, H, H^2, H^3\}$, since in any dimension
$H^4=I$. This means that the eigenvalues of the Hadamard operator
are restricted to the set $\{1, -1, i, -i\}. $ Again this group
is Abellian and all its irreducible representations are one dimensional. Taking as the
representations of $H$, its defining representation which is
reducible, we find from (\ref{remark1}), the following
\begin{equation}\label{Hada}
  H^{-1} A^{(\m)} H=\mu A^{(\m)},
\end{equation}
where $\mu\in \{1,-1,i,-i\}$. The solutions for $A^{(\m)}$ are
obtained in the same way as before from the eigenvectors of the
operator $H$. The above considerations apply for any dimension,
for the three dimensional case, we have to note that the
eigenvalues of the three dimensional Hadamard operator
\begin{equation}\label{Hadamard}
H=\frac{1}{\sqrt{3}}\left(\begin{array}{ccc} 1 & 1 & 1 \\ 1 &
\omega & \omega^2 \\ 1 & \omega^2 & \omega
\end{array}\right),
\end{equation}
are confined to the subset $\{1,-1,i\}$. This can be verified
either by explicit calculations of the eigenvalues or by noting
that $tr(H)=i$ and $tr(H^2)=1$.\\

Let us denote the orthonormal set of eigenvectors by $|\eta_1\ra,
\ |\eta_{-1}\ra$ and $|\eta_i\ra$ respectively. A simple
calculation shows that their un-normalized form are as follows

\begin{eqnarray}\label{eigH}
  |\eta_{_1}\ra=\left(\begin{array}{c}1+\sqrt{3}\\ 1 \\
  1\end{array}\right),\h  |\eta_{_1}\ra=\left(\begin{array}{c}1-\sqrt{3}\\ 1 \\
  1\end{array}\right),\h |\eta_{_i}\ra=\left(\begin{array}{c}0\\ -1 \\
  1\end{array}\right).
\end{eqnarray}
With a judicious choice of the labeling of free parameters, the
solution of (\ref{Hada}) will be given by
\begin{eqnarray}\label{hadasol}
  A^{^{(1)}} &=&a_{11}|\eta_{_1}\ra\la \eta_{_1}|+a_{22}|\eta_{_{-1}}\ra\la
  \eta_{_{-1}}|+a_{33}|\eta_{_{i}}\ra\la \eta_{_{i}}|, \cr
 A^{^{(-1)}} &=&a_{21}|\eta_{_1}\ra\la \eta_{_{-1}}|+a_{12}|\eta_{_{-1}}\ra\la
  \eta_{_{1}}|, \cr
   A^{^{(-i)}} &=&a_{31}|\eta_{_1}\ra\la \eta_{_i}|+a_{23}|\eta_{_{i}}\ra\la
  \eta_{_{-1}}|, \cr
A^{^{(i)}} &=&a_{13}|\eta_{_i}\ra\la
\eta_{_1}|+a_{32}|\eta_{_{-1}}\ra\la
  \eta_{_{i}}|.
\end{eqnarray}
Using the orthonormal property of the eigenvectors, and defining
the vectors ${\bf a}_i:=(a_{i1}, a_{i2}, a_{i3})$ and $\tilde {\bf
a}_i:=(a_{1i}, a_{2i}, a_{3i})$, we find that

\begin{equation}\label{Hadtrace}
  \sum {A^{{(\m)}}}^\dagger A^{{(\m)}}=
  |{\bf a}_1|^2|\eta_{_1}\ra\la \eta_{_1}|+|{\bf a}_2|^2|\eta_{_{-1}}\ra\la \eta_{_{-1}}|+|{\bf a}_3|^2|\eta_{_i}\ra\la
  \eta_{_i}|
\end{equation}
and

\begin{equation}\label{Hadtrace}
  \sum A^{{(\m)}}{A^{{(\m)}}}^\dagger =
  |\tilde{\bf a}_1|^2|\eta_{_1}\ra\la \eta_{_1}|+|\tilde{\bf a}_2|^2|\eta_{_{-1}}\ra\la \eta_{_{-1}}|+|\tilde{\bf a}_3|^2|\eta_{_i}\ra\la
  \eta_{_i}|.
\end{equation}
Therefore the CP map will be trace-preserving if the vectors
${\bf a}_i$ are of unit length and will be unital if the vectors
$\tilde{\bf a}_i$ are of unit length.\\

Certainly one can study other examples of cyclic groups, for
example a group generated by one single element which swaps the
basis states $|1\ra$ and $|2\ra$ or a group which is generated by
a single discrete phase operator $|1\ra\lo |1\ra, |2\ra\lo
e^{\frac{2\pi i k_1}{n}}|2\ra, |3\ra\lo e^{\frac{2\pi i
k_2}{n}}|3\ra$. However we now consider a non-Abelian discrete
group, the simplest of which is the generalized Pauli group.

\subsection{The Non-Abelian Case: Pauli and Permutation Groups}

\textbf{i) Pauli Group} As the first example in this class, we
consider the generalized Pauli group, whose
elements consists of generalized Pauli operators  $\{X_{mn}=X^mZ^n =
\sum_{j=0}^{d-1} \omega^{jn}|j+m\ra\la j|\}$, where $\omega = e^{\frac{2 \pi i}{d}}$ and $X$, $Z$
are the generalized $\sigma_x$ and $\sigma_z$ operators with
$X|j\ra=|j+1\ra$ and $Z|j\ra=\omega^j|j\ra$. Due to the simple
commutation
$$X_{kl}X_{mn}=\omega^{lm-kn}X_{mn}X_{kl},$$ the collection of
all the Pauli operators and their multiples of discrete powers of
$\omega$ make a group, which is called Pauli group. From the
above relation one easily obtains
$$X_{mn}^{\dagger}X_{kl}^{\dagger}=\omega^{-(lm-kn)}X_{kl}^{\dagger}X_{mn}^{\dagger},$$
from which we find that any channel of the following form, i.e. a Pauli channel,
\begin{equation}\label{pauli}
  {\cal E}_{p}(\rho):=\sum_{i,j}p_{ij}X_{ij}\rho X_{ij}^{\dagger},
\end{equation}
is covariant under the generalized Pauli group.\\

To find the symmetry properties of a Pauli channel, consider the
case where the channel is symmetric under one Pauli operator
$X_{mn}$, i.e. ${\cal E}_p(\rho)={\cal E}_p(X_{mn}\rho X_{mn}^{\dagger})$.
Using the Kraus decomposition of this channel (\ref{pauli}) and the relation
$X_{kl}X_{mn}=\omega^{ml}X_{k+m,l+n}$, we find that the channel
will be symmetric provided that the following relations hold
among the error probabilities,
\begin{equation}\label{paulisym}
p_{kl}=p_{k+m,l+n}\h \forall (k,l).
\end{equation}
Such a channel is symmetric under the action of a subgroup
$H\subset G$ of the Pauli group, generated by $X_{mn}$. Let this
subgroup be of size $r$. According to Lagrange's theorem, $r$
divides the size of the group $Z_d\times Z_d$. Equation
(\ref{paulisym}) shows that the error probabilities are constant
in each co-set of the subgroup, so in total there are $d^2/r-1$
independent parameters for the channel. For qutrits, since $d=3$
is a prime number, it is readily verified that the symmetry under
any subgroup $\la X_{mn}\ra$ generated by one single operator
$(m,n)\ne (0,0)$, reduces the number of parameters from $8$ to
$2$. For example a channel which is symmetric under $\la X_{01}=Z
\ra$ has the following form
\begin{equation}\label{symmpauli01}
  {\cal E}_{01}(\rho)=\sum_{i,j=0}^2 p_i X_{ij}\rho X_{ij}^{\dagger}.
\end{equation}
This channel is also covariant under Pauli group. The symmetry property, i.e. ${\cal
E}(\rho)={\cal E}(Z\rho Z^{\dagger})$, implies that the minimum
output entropy states are the computational basis vectors,
$|0\ra, \ |1\ra$, and $|2\ra$, each with the same output entropy
given by
\begin{equation}\label{hhh}
  H(p)=-(p_0\log p_0+p_1\log p_1+p_2\log p_2).
\end{equation}
This leads to the one-shot capacity $C=\log_{3}3 - H(p).$  \\

Another example is a channel which is both Pauli covariant and symmetric
under $\la X_{10}=X\ra$
\begin{equation}\label{symmpauli10}
  {\cal E}_{10}(\rho)=\sum_{i,j=0}^2 p_j X_{ij}\rho
  X_{ij}^{\dagger}.
\end{equation}
Since the channel is symmetric under the action of $X$,  the
minimum output entropy states are the $X$-invariant states, i.e.
eigenstates of $X$, which are $|\xi_n\ra:=\frac{1}{\sqrt{3}}\sum_j
\omega^{nj}|j\ra$ ($\omega^3=1$), giving the same output entropy
and the same one-shot capacity as in the previous example
(\ref{hhh} ).\\

Finally a channel which is Pauli covariant and symmetric under $\la
X_{11}=XZ\ra$ is as follows:
\begin{equation}\label{symmpauli11}
  {\cal E}_{11}(\rho)=\sum_{i,j=0}^2 p_j X_{i,i+j}\rho X_{i,i+j}^{\dagger}.
\end{equation}
Similar arguments as before show that the one-shot capacity of
this channel is also given by (\ref{hhh}).\\
\\
\textbf{ii) Permutation Group} As another example of a
non-Abelian discrete group, consider the permutation group $S_3$
whose action on the input qutrit state $a|0\ra+b|1\ra+c|2\ra$ is
generated by two unitary operators, which we denote by $\sigma_1$
and $\sigma_2$. Here $\sigma_1$ interchanges only the
computational states $|0\ra$ and $|1\ra$, while $\sigma_2$
interchanges the basis states $|1\ra$ and $|2\ra$. We take the
representations $D^{^{(1)}}$ and $D^{^{(2)}}$ to coincide with
this defining representation. Therefore we have

\begin{equation}\label{perm}
D^{^{(1)}}(\sigma_1)=D^{^{(2)}}(\sigma_1) =
\left(\begin{array}{ccc} 0 & 1 & 0 \\ 1 & 0 & 0 \\ 0 & 0 & 1
\end{array}\right), \h D^{^{(1)}}(\sigma_2)=D^{^{(2)}}(\sigma_2) =
\left(\begin{array}{ccc} 1 & 0 & 0 \\ 0 & 0 & 1 \\ 0 & 1 & 0
\end{array}\right).
\end{equation}
The elements of the permutation group $S_3$ are given as $S_3 =
\{e, \sigma_1, \sigma_2, \sigma_1\sigma_2, \sigma_2\sigma_1,
\sigma_1\sigma_2\sigma_1\}$, where $e$ is the identity element
and the relations $\sigma_1^2=\sigma_2^2=e$ and
$\sigma_1\sigma_2\sigma_1=\sigma_2\sigma_1\sigma_2$ hold.\\

The group $S_3$ has three inequivalent irreducible
representations. These are two 1-dimensional ones, which we
denote by $\Omega^{^{(1)}}$ and $\Omega^{^{(1')}}$ and a
2-dimensional one which we denote by $\Omega^{^{(2)}}$. These are

\begin{equation}\label{}
\Omega^{^{(1)}}(\sigma_1)=\Omega^{^{(1)}}(\sigma_2)=1,
\end{equation}

\begin{equation}\label{}
\Omega^{^{(1')}}(\sigma_1)=\Omega^{^{(1')}}(\sigma_2)=-1
\end{equation}
and
\begin{equation}\label{}
\Omega^{^{(2)}}(\sigma_1)=\left(\begin{array}{cc} 1 & 0  \\ 0 &
-1 \end{array}\right), \hh
\Omega^{^{(2)}}(\sigma_2)=\frac{1}{2}\left(\begin{array}{cc} -1 & \sqrt{3}  \\
\sqrt{3} & 1 \end{array}\right).
\end{equation}

We consider these representations separately and then combine the
results to find a channel which is covariant with respect to
permutation group. Dropping for simplicity the symbols
$D^{^{(1)}}$ and $D^{^{(2)}}$ in the basic equation (\ref{covfinal}), we have
for the representation $\Omega^{^{(1)}}$ one single Kraus
operator satisfying the following two equations

\begin{equation}\label{}
  \sigma_1 A \sigma_1 = A, \h \sigma_2 A \sigma_2 = A,
\end{equation}
the solution of which is given by
\begin{equation}\label{A}
  A = \left(\begin{array}{ccc} a & b & b \\ b & a & b \\ b & b &
  a\end{array}\right).
\end{equation}\\
For the representation $\Omega^{^{(1')}}$ the single Kraus
operator $B$ should satisfy the following two equations

\begin{equation}\label{}
  \sigma_1 B \sigma_1 = -B, \h \sigma_2 B \sigma_2 = -B,
\end{equation}
with the solution given by
\begin{equation}\label{B}
  B = \left(\begin{array}{ccc} 0 & c & -c \\ -c & 0 & c \\ c & -c &
  0\end{array}\right).
\end{equation}
Finally for the representation $\Omega^{^{(2)}}$, we have two
Kraus operators which should satisfy the following equations

\begin{equation}\label{C2}
\sigma_1C_1 \sigma_1 = C_1, \h \sigma_1 C_2 \sigma_1 = - C_2
\end{equation}
and

\begin{equation}\label{C2}
\sigma_2C_1 \sigma_2 = \frac{-1}{2}C_1+\frac{\sqrt{3}}{2}C_2, \h
\sigma_2 C_2 \sigma_2 = \frac{\sqrt{3}}{2}C_1+\frac{1}{2}C_2,
\end{equation}
the solution of which is
\begin{equation}\label{C1C2}
  C_1 = \left(\begin{array}{ccc} d&-e-f& e \\ -e-f& d& e \\ f& f&
  -2d\end{array}\right), \h C_2 = \frac{1}{\sqrt{3}}\left(\begin{array}{ccc} 3d&e-f&-e-2f \\ f-e& -3d& e+2f \\ -2e-f& 2e+f&
  0\end{array}\right).
\end{equation}\\
\\
Any CP map of the form
\begin{equation}\label{S3map}
  {\cal E}(\rho) = A\rho A^{\dagger}+B\rho B^{\dagger}+C_1\rho
  C_1^{\dagger}+C_2\rho C_2^{\dagger},
\end{equation}
is covariant with respect to the permutation group $S_3$. The
above CP map has $6$ free parameters. To put the additional
condition of trace-preserving CP map, we have to solve the
equation
\begin{equation}\label{S3tp}
  A^{\dagger}A+B^{\dagger}B+C_1^{\dagger}C_1+C_2^{\dagger}C_2=I.
\end{equation}
This condition constrains the parameters to a smaller manifold.\\

For this channel to be symmetric under permutation group, we have
to solve equations (\ref{symkrauscond}). For the representations $\Omega^{^{(1)}}$
this takes the form

\begin{equation}\label{permsym1}
 A \sigma_1 = A, \h A\sigma_2 = A,
\end{equation}
its solution is given by
$$A= \left(\begin{array}{ccc} a & a & a \\ b& b & b \\ c & c
& c
\end{array}\right),$$
where $a,b$ and $c$ are free parameters. For the representations
$\Omega^{^{(1')}}$ this takes the form

\begin{equation}\label{permsym2}
 B \sigma_1 = -B, \h B\sigma_2 = -B,
\end{equation}
whose solution is $B=0$. Finally for the representation
$\Omega^{^{(2)}}$, the equations are

\begin{equation}\label{permsym3}
  C_1 \sigma_1 = C_1, \h  C_2\sigma_1 = - C_2
\end{equation}
and

\begin{equation}\label{permsym3}
  C_1 \sigma_2 = \frac{1}{2}(-C_1 + \sqrt{3} C_2), \h C_2 \sigma_2 = \frac{1}{2}(\sqrt{3}C_1 +  C_2),
\end{equation}
the solution of which is

\begin{equation}\label{C1C2sym}
C_1 = \left(\begin{array}{ccc} d & d & -2d \\ e& e & -2e \\ f & f
& -2f
\end{array}\right), \h C_2 = \sqrt{3}\left(\begin{array}{ccc} d & -d & 0 \\ e& -e & 0 \\ f & -f
& 0
\end{array}\right).
\end{equation}
A simple calculation shows that the following completely positive
map which is symmetric under permutation group,

$${\cal E}_S(\rho):=A\rho A^{\dagger}+C_1^{\dagger} \rho C_1 + C_2^{\dagger} \rho C_2,$$
will also be trace-preserving provided that the parameters
satisfy the following conditions:
\begin{equation}\label{symmrelation}
  a=b=c=\frac{1}{3},\h  |d|^2+|d|^2+|f|^2=\frac{1}{6}.\
\end{equation}

Clearly many special cases in this class with simple solutions can be
considered. It is now desirable to leave the examples of
discrete transformation groups and continue with the investigation of
examples from continuous groups.

\section{Continuous Groups}\label{contexam}
We consider three continuous groups acting on qutrit states,
namely $U(1)$, $U(1)\times U(1)$ and $SU(3)$. The first two are
Abelian and the third one is non-Abelian.

\subsection{The U(1) group}

As our first example of a continuous group of transformations,
let us consider a group of phase shift operators, whose action on
any qutrit state is defined as
$g(\theta)(a|0\ra+b|1\ra+c|2\ra=a|0\ra+b|1\ra+ce^{i\theta}|2\ra).$
This group is isomorphic to $U(1)$ whose irreducible
representations are all one dimensional and are labeled by a real
number $\a\in [0,2\pi]$, i.e.
$\Omega^{(\a)}(g=e^{i\theta})=e^{i\alpha \theta}$. Taking
$D^{(1)}(g)=D^{(2)}(g)=g=diagonal (1,1,e^{i\theta})$, we have to
solve the following equation

\begin{equation}\label{u1}
  g^{-1}A^{\a}g=e^{i\a\theta} A^{\a},
\end{equation}
whose solution depend on the value of $\a$. The only
representations (i.e. values of $\a$) which yield non-zero
solutions are found to be
\begin{equation}\label{AU1}
  A^{(0)} = \left(\begin{array}{cc} B & {\bf 0} \\ {\bf 0}  & a
  \end{array}\right),
\end{equation}
where $B$ is an arbitrary two-dimensional matrix,
\begin{equation}\label{AU1}
  A^{(1)} = \left(\begin{array}{ccc} 0 & 0 & 0 \\ 0 & 0 & 0 \\ b & c & 0
  \end{array}\right)
\end{equation}
and

\begin{equation}\label{AU1}
  A^{(-1)} = \left(\begin{array}{ccc} 0 & 0 & d \\ 0 & 0 & e \\ 0 & 0 & 0
  \end{array}\right).
\end{equation}
The covariant channel under these $U(1)$ transformations will be
of the form

\begin{equation}\label{u1channel}
  {\cal E}(\rho) = A^{(0)}\rho {A^{(0)}}^{\dagger}+ A^{(1)}\rho {A^{(1)}}^{\dagger}+A^{(-1)}\rho {A^{(-1)}}^{\dagger},
\end{equation}
where for trace-preserving property, we should have
\begin{equation}\label{u1trace}
  B^{\dagger}B + \left(\begin{array}{cc} |b|^2 & 0 \\ 0 &
  |c|^2\end{array}\right)=I, \h |a|^2+|d|^2+|e|^2=1.
\end{equation}\\

\subsection{The $U(1)\times U(1)$ group}

Another interesting continuous group which is Abelian has the
following action on qutrits,
$g(\theta_1,\theta_2)$ $(a|0\ra+b|1\ra+c|2\ra)=a|0\ra+e^{i\theta_1}|1\ra+e^{i\theta_2}|2\ra$.
This group is isomorphic to $U(1)\times U(1)$ whose irreducible
representations are defined by two real numbers,
$\Omega^{(\a_1,\a_2)}(g)=e^{i\alpha_1\theta_1+\alpha_2\theta_2}.$
Proceeding along the same lines as before we find that only for a
limited number of representations there are non-zero solutions,
and these solutions are

\begin{equation}\label{u1u1first}
  A^{(0,0)}=a_0|0\ra\la 0|+a_1|1\ra\la 1| + a_2|2\ra\la 2|
\end{equation}
and
\begin{equation}\label{u1u1second}
  A^{(i,-j)}=|i\ra\la j|,\ \ \ (i,j)\ne (0,0).
\end{equation}
The channel will be of the form

\begin{equation}\label{Eu1u1}
  {\cal E}(\rho) = A^{(0,0)}\rho {A^{(0,0)}}^{\dagger} + \sum_{(i,j)\ne
  (0,0)}p_{ij}|i\ra\la j|\rho|j\ra\la i|.
\end{equation}
It is readily found that this CP map will be trace preserving
provided that the following condition holds:
$$|a_j|^2+\sum_{i=0}^2 p_{ij}=1,\ \ \ \ \ j=0,1,2.$$
These considerations can easily be generalized to the $d-$
dimensional case.

\subsection{The SU(3) group}\label{Su3sub}

When considering non-Abellian continuous groups, we can resort to
the infinitesimal generators, i.e. the elements of the Lie
algebra of the group. These relations render all the relations
linear and easy to solve. Let $G$ be a continuous group of
transformations on the input state. The local coordinates and the
infinitesimal generators of $G$ are denoted respectively by
$\theta_n$ and $T_n$, i.e. $g=e^{i\sum_n \theta_n T_n}$.  Any
representation of the Lie algebra induces a representation of the
Lie group. In this case we have
\begin{equation}\label{groupinfinitesimal}
  D^{^{(1)}}(g) = e^{i\theta_n D^{^{(1)}}(T_n)},\h  D^{^{(2)}}(g) = e^{i\theta_n
  D^{^{(2)}}(T_n)},\h \Omega(g)=e^{i\theta_n \Omega ({T}_n)}.
\end{equation}
In terms of Lie algebra generators, condition (\ref{covfinal}) now reads
\begin{equation}\label{infinitesimal}
  A_a D^{^{(1)}}(T_n)- D^{^{(2)}}(T_n) A_a = \sum_b {\Omega({T_n})}_{ab} A_b.
\end{equation}
For qutrit channels $D^{^{(1)}}$ and $D^{^{(2)}}$ are three
dimensional representations of the group, and the dimension of
$\Omega$ determines the number of Kraus operators. \\

The $SU(3)$ group has also a natural action on a qutrit and it is
desirable to study qutrit channels which are covariant or
symmetric under this group. The interesting point about this
groups is that there are two inequivalent irreducible
3-dimensional representations, denoted as $3$ (or quark) and
$\overline{3}$ (or anti-quark) \cite{georgi} and there is a possibility that the
channel be covariant under different input and output
representations. It is interesting to investigate this possibility.
To explore fully the covariance property of a qutrit channel with
respect to this group, we proceed as before by treating all the
possible irreducible representations for the matrix $\Omega$. For
any given channel on a qutrit, the maximum number of Kraus
operators can always be reduced to $9$, which is the square of
the dimension of the Hilbert space. There are a finite number of
irreducible representations of $su(3)$ with dimension less than
$9$. So once we analyze these representations and the
corresponding covariant channels, we will be able to construct all
the other channels, simply be taking the convex combination of
such covariant channels.\\

The basic facts about the Lie algebra $su(3)$ and it's irreducible
representations are collected in the appendix. The material
collected in this appendix is essential for the method we use for
solving the basic equation (\ref{infinitesimal}). In order to solve these equation,
we use the vectorized form of the Kraus operators $A_a$. That is
we write a matrix $A=\sum_{i,j}A_{i,j}|i\ra\la j|\in M_d$ as a
vector $|A\ra = \sum_{i,j}A_{i,j}|i,j\ra\in C^d\otimes C^d$. In
this notation, the following product of matrices take the
following forms

\begin{equation}\label{matrix}
|BA\ra = (B\otimes I)|A\ra,\h |AB\ra = (I\otimes B^T)|A\ra.
\end{equation}
With these notations, and by using the definition of the
conjugate representation, namely $\overline{D}(x) = - [D(x)]^T$,
equation (\ref{infinitesimal}) will transform to

\begin{equation}\label{transform}
  (I\otimes \overline{D^{^{(1)}}}(T_m) +  D^{^{(2)}}(T_m)\otimes I)|A_a\ra =[\overline{\Omega}(T_m)]_{ba}|A_b\ra.
\end{equation}
This equation not only gives us the explicit solutions of the
Kraus operators, but also it readily gives the condition under
which non-zero solutions exists. Since the operator in the left
hand side is nothing but the representation of $T_m$ in the
tensor product of $\overline{D^{^{(1)}}}$ and $D^{^{(2)}}$ \cite{groupthe}, we
conclude that nonzero solutions of (\ref{infinitesimal}) exist only if the
representation $\overline{\Omega}$ is contained in the
decomposition of $\overline{D^{^{(1)}}} \otimes D^{^{(2)}}$ or by
conjugating both sides, if

\begin{equation}\label{condition}
\Omega \subset D^{^{(1)}}\otimes  \overline{D^{^{(2)}}}.
\end{equation}

So if $D^{^{(1)}}$ and $D^{^{(2)}}$ are irreducible
representations, then in view of this condition and the rules (\ref{tensor})
for decomposition of tensor products of representations of
$su(3)$ \cite{georgi}, we find that equation (\ref{condition}) allows only the solutions
collected in table 1. \\

\begin{table}\label{jadval}
\centering
\begin{tabular}{ccc}
\hline
\ \ \ $D^{^{(1)}}$ &  \ \ \ $D^{^{(2)}}$ & $\Omega$ \\
\hline 3 & 3 & 8 \ \ or\ \ 1 \\
3 & $\overline{3}$ & 6 \ \ or\ \ $\overline{3}$ \\
$\overline{3}$ & 3 & $\overline{6}$\ \ or\ \ 3 \\
$\overline{3}$ & $\overline{3}$ & 8 \ \ or\ \ 1 \\
\hline
\end{tabular}
\caption{The allowed representations for solving equation (\ref{infinitesimal}).}
\end{table}

The first row of table (1), gives us two solutions with 8 and 1
Kraus operators respectively, whose vectorized forms transform
under the representations $8$ and $1$ of $su(3)$. From the
construction given in the appendix, these vectors are given by

\begin{equation}\label{8vec}
  |A^{8}_{ij}\ra=|\mu_i\ra|\overline{\mu}_j\ra -\frac{1}{3}\delta_{ij}(\sum_{k=1}^3|\mu_k\ra|\overline{\m}_k\ra)
  \end{equation}
and

\begin{equation}\label{1vec}
  |A^{1}_{ij}\ra=\delta_{ij}(\sum_{k=1}^3|\mu_k\ra|\overline{\m}_k\ra),
  \end{equation}
which gives the Kraus operators

\begin{equation}\label{8vec2}
  A^{8}_{ij}=|\mu_i\ra\la \overline{\mu}_j|-\frac{1}{3}\delta_{ij}(\sum_{k=1}^3|\mu_k\ra\la
  \overline{\m}_k|)=E_{ij}-\frac{1}{3}\delta_{ij}I
  \end{equation}
and

\begin{equation}\label{1vec2}
  A^{1}_{ij}=\delta_{ij}(\sum_{k=1}^3|\mu_k\ra\la
  \overline{\m}_k|)=\delta_{ij}I.
  \end{equation}
Thus we obtain two trace preserving maps covariant under $3$ and $3$ in the form
\begin{equation}\label{E8}
  {\cal E}^{8}(\rho) = \frac{1}{2} \sum_{ij}(E_{ij}-\frac{1}{3}\delta_{ij}I)\rho
  (E_{ji}-\frac{1}{3}\delta_{ij}I)=\frac{1}{2} (tr(\rho) I -\rho)
\end{equation}
and
\begin{equation}\label{E8}
  {\cal E}^{1}(\rho) = \rho.
\end{equation}
A similar reasoning from the last row of table (1) gives the same
set of Kraus operators and the same map as above, which is also
covariant with
respect to $\overline{3}$ and $\overline{3}$.\\

The convex combination of these two maps has the same covariance property and is given by

\begin{equation}\label{cov8}
  {\cal E}(\rho) = \frac{1}{2} [p\ tr(\rho) I + (2-3p)\rho],
\end{equation}
which is a one parameter trace-preserving and unital channel with $0 \leq p \leq 1$.\\

The capacity $C^{(1)}$ is easily found for this channel. Since
the action of $su(3)$ is transitive on all qutrits, the output
entropy of all pure states are the same, so we should only find
an ensemble of pure states that maximizes the first term of
Holevo quantity. The ensemble $\{ |0>, |1>, |2> \}$ with uniform
probability distribution is the intended ensemble. A simple
calculation leads to
\begin{equation}
C^{(1)} = \log(3) +(1-p) \log(1-p) + p \log(\frac{p}{2}).
\end{equation}

Consider now the second row of the table (1). There are two kinds
of map here, one with $6$  and the other with $3$ Kraus
operators, both of which are covariant with respect to the
representations $3$ and $\overline{3}$. From the relations in the
appendix , the vectors of $6$ are given by

\begin{equation}\label{6vec}
  |A^{6}_{ij}\ra=|\mu_i\ra|\mu_j\ra+|\mu_j\ra|\mu_i\ra,
\end{equation}
and those of $\overline{3}$ are given by

\begin{equation}\label{6vec}
  |A^{\overline{3}}_{ij}\ra=|\mu_i\ra|\mu_j\ra-|\mu_j\ra|\mu_i\ra,
\end{equation}
leading to the Kraus operators

\begin{equation}\label{6vec2}
  A^{6}_{ij}=E_{ij}+E_{ji},
\end{equation}
and those of $\overline{3}$ are given by

\begin{equation}\label{6vec}
  A^{\overline{3}}_{ij}=E_{ij}-E_{ji}.
\end{equation}
The corresponding positive trace preserving covariant maps are given by

\begin{equation}\label{E6}
  {\cal E}^{6}(\rho) = \frac{1}{8}\sum_{ij}(E_{ij}+E_{ji})\rho
  (E_{ji}+E_{ij})=  \frac{1}{4} (tr(\rho) I + \rho^T )
\end{equation}
and

\begin{equation}\label{E3}
  {\cal E}^{\overline{3}}(\rho) =\frac{1}{4} \sum_{ij}[i(E_{ij}-E_{ji})]\rho
  [-i(E_{ji}-E_{ij})]= \frac{1}{2} ( tr(\rho) I - \rho^T ).
\end{equation}
From the third row of table (1) we see that the maps corresponding
to $\overline{6}$ and $3$ are the same as (\ref{E6}) and
(\ref{E3}) and hence these two maps are also covariant with
respect to the representations
$\overline{3}$ and $3$.\\

Finally the convex combination of these two channels has the same
covariance property and will be a CPT map by $ 0 \le p \le 1,$

\begin{equation}\label{cov6}
  {\cal E}(\rho) = \frac{1}{4} [(2-p)\ tr(\rho) I + (3p-2)\ \rho^T].
\end{equation}
Following the same reasoning as in the previous case, we find the
one-shot capacity to be
\begin{equation}
C^{(1)} = \log(3) + \frac{p}{2} \log(\frac{p}{2}) + \frac{2-p}{2}
\log(\frac{2-p}{4}).
\end{equation}\\

\section{Summary and Outlook}
We have studied the problem of characterizing qutrit channels
from a different point of view than previously done, namely we have
focused on the covariance and symmetry properties of such
channels to categorize qutrit channels. By using the Kraus representation of such maps, we have
developed a formalism which turns the investigation of such
channels, not only for qutrits but for any channel and for any
transformation group into a systematic problem in the representation
theory of the group and its algebra. Although our examples are
mainly for the qutrit channels, to comply with the main theme of
our work, this formalism has much wider application and we hope
that other authors will apply this method for study of a much
larger class of channels.\\

Needless to say, this is only a first step toward understanding
the space of completely positive maps on three dimensional
matrices. There is a long road ahead to gain a complete
understanding of this space.\\

{\textbf{Acknowledgements}} We would like to thank  S. Alipour ,
S. Baghbanzadeh, M. R. Koochakie,   A. T. Rezakhani and M. H.
Zare for valuable comments and interesting discussions.

\vskip 1cm

 \section{Appendix: Some basic facts about $su(3)$ and its representations}
\begin{figure}
  \centering
  \epsfig{file=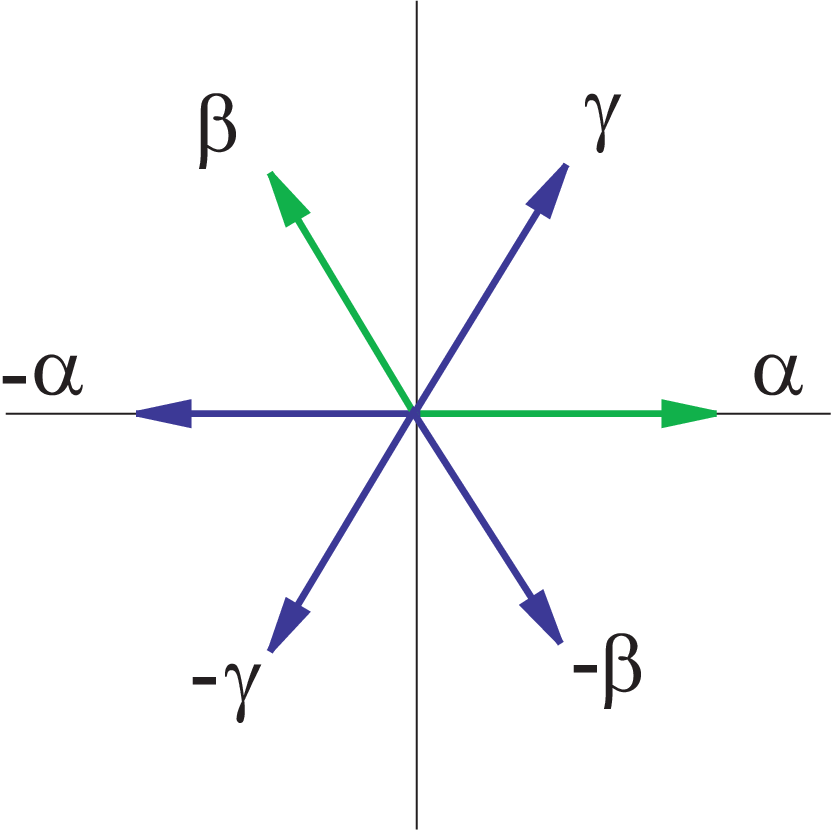,width=4.5cm}
  \caption{(Color online) The root diagram of $su(3)$.}
  \label{su3roots}
\end{figure}

For ease of reference, we collect here some basic facts about the
$su(3)$ algebra and its representations \cite{groupthe,georgi}.
The algebra $su(3)$ is a rank-2 algebra with two commuting
elements (i.e. basis of Cartan subalgebra) $H_1$ and $H_2$
$[H_1,H_2]=0$. The other generators of $su(3)$ can be organized
in such a way to be common eigenvectors of these two generators
under commutation (or adjoint action in more mathematical term),
that is:
\begin{equation}\label{su3relation}
  [H_1, E_{\bf \a}]=\a_1 E_{\bf \a}, \h  [H_2, E_{\bf \a}]=\a_2 E_{\bf \a},
\end{equation}
where there are six two dimensional vectors ${\bf a}$ (called
roots) and correspondingly six other generators. The roots of
$su(3)$, like any other Lie algebra, have a very rigid structure,
reflecting the rigid structure of the commutation relations of
the algebra. Usually they are organized in a diagram called the
root diagram. The roots $\a, \b$ and $\gamma$ act as raising operators in any
representation, while $-\a, -\b$, and $-\gamma$ act as lowering
operators. Figure (\ref{su3roots}) shows the root diagram of $su(3)$.\\

Apart from the trivial one dimensional representation, where all
the generators are assigned by the number $0$, there are a countably
infinite number of unitary irreducible representations of $su(3)$.
 For any Lie algebra and any
unitary representation of it say, $D$, there is a complex
conjugate representation $\overline{D}$, where
$\overline{D}(T)=[-D(T)]^T$. To see this one needs to invoke the
fact that in a unitary representation of a group, the generators
are represented by Hermitian matrices, so if $D(T_a)$ satisfy the
commutation relations of an algebra, so do $\overline{D}(T_a).$
As in the simpler case of $su(2)$, any representation of $su(3)$
is specified by its weights, that is the common eigenvalues of its
vectors $|{\bf \mu}\ra $ for the commuting operators $H_1$ and
$H_2$;
\begin{equation}\label{wei}
  H_1|{\bf \m}\ra = \mu_1 |{\bf \mu}\ra, \h H_2|{\bf \m}\ra = \mu_2 |{\bf \mu}\ra.
\end{equation}

\begin{figure}
  \centering
  \epsfig{file=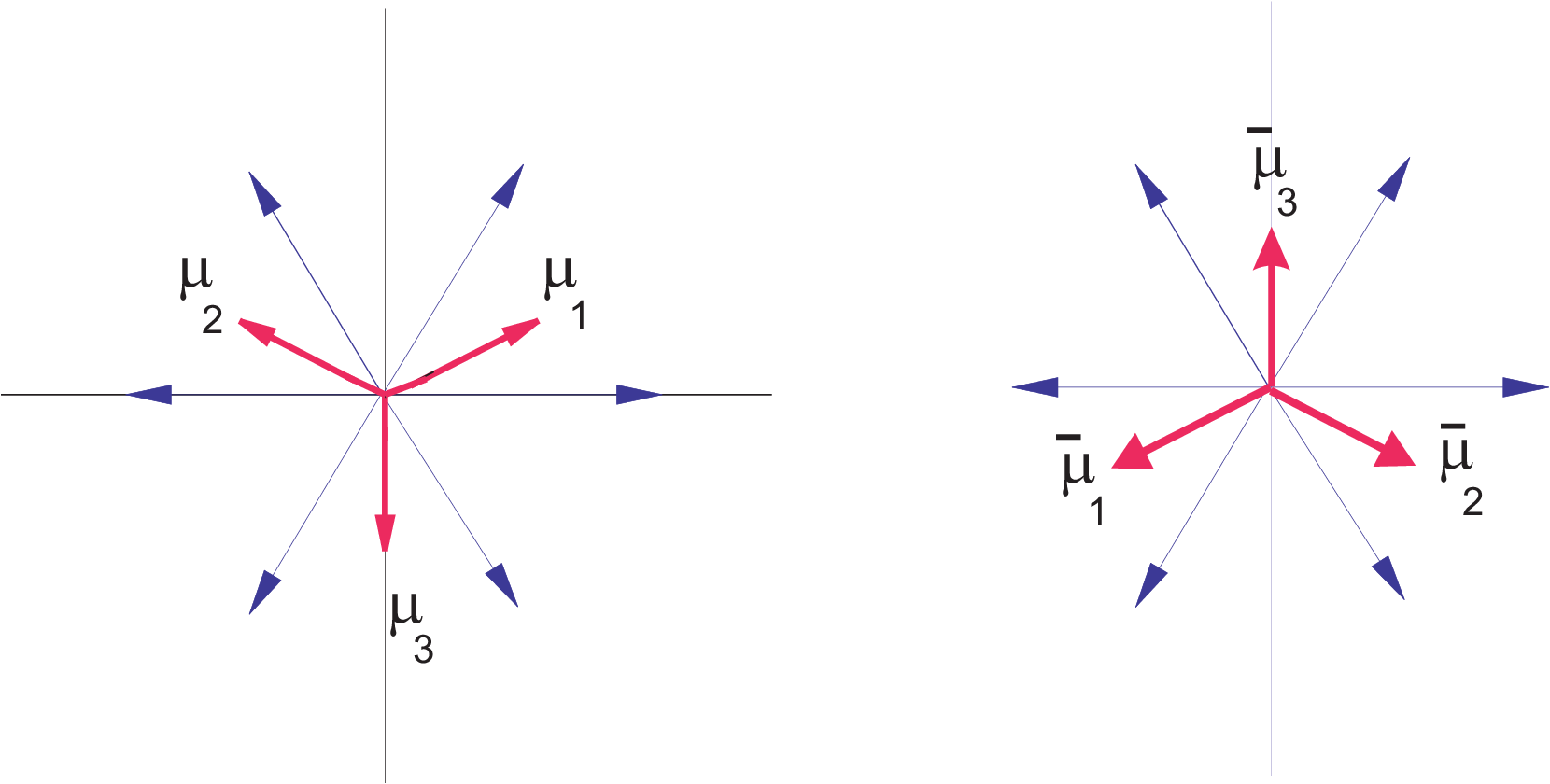,width=10cm}
  \caption{(Color online) The weight diagrams of the representations $3$ (left) and $\overline{3}$
  (right).}
    \label{su3antiquark}
\end{figure}

There are two three dimensional representations which we denote
simply by $3$ and $\overline{3}$. Their weight diagrams are shown
in figure (\ref{su3antiquark}).
Note that the weight diagram of $\overline{3}$ is obtained from
that of $3$ by a reflection through the origin. The basis vectors
of the $3$ representation are:

\begin{equation}\label{basis3}
  |\mu_1\ra=|\frac{1}{2},\frac{1}{2\sqrt{3}}\ra=\left(\begin{array}{c}1\\ 0 \\ 0
  \end{array}\right),\ \ \  |\mu_2\ra=|\frac{-1}{2},\frac{1}{2\sqrt{3}}\ra=\left(\begin{array}{c}0\\ 1 \\ 0
  \end{array}\right),\ \ \ |\mu_3\ra=|0,\frac{-1}{\sqrt{3}}\ra\left(\begin{array}{c}0\\ 0 \\ 1
  \end{array}\right).
\end{equation}
In such a representation the Cartan matrices are represented by

\begin{equation}\label{H3Rep}
H_1=\frac{1}{2}\left(\begin{array}{ccc}1 & & \\ & -1 & \\ & & 0
\end{array}\right), \h  H_2=\frac{1}{2\sqrt{3}}\left(\begin{array}{ccc}1 & & \\ & 1 & \\ & & -2
\end{array}\right)
\end{equation}
Similarly the basis vectors of the representation $\overline{3}$
are

\begin{equation}\label{basis3}
  |\overline{\mu}_1\ra=|-\frac{1}{2}, -\frac{1}{2\sqrt{3}}\ra=\left(\begin{array}{c}1\\ 0 \\ 0
  \end{array}\right),\ \ \  |\overline{\mu}_2\ra=|\frac{1}{2},\frac{-1}{2\sqrt{3}}\ra=\left(\begin{array}{c}0\\ 1 \\ 0
  \end{array}\right),\ \ \ |\overline{\mu}_3\ra=|0,\frac{1}{\sqrt{3}}\ra\left(\begin{array}{c}0\\ 0 \\ 1
  \end{array}\right).
\end{equation}
In this representation, the Cartan matrices are represented by
\begin{equation}\label{basis3}
H_1=\frac{1}{2}\left(\begin{array}{ccc}-1 & & \\ & 1 & \\ & & 0
\end{array}\right), \h  H_2=\frac{1}{2\sqrt{3}}\left(\begin{array}{ccc}-1 & & \\ & -1 & \\ & & 2
\end{array}\right).
\end{equation}

Other representations of small dimensions which we need for our
discussions are $6$, $\overline{6}$, and $8$, where again the
numbers denote dimensions of the representations. Note that a
representation like $8$ is self-conjugate (real). The weight
diagram of this representation is symmetric under reflection
through the origin. Like $su(2)$, higher dimensional
representations of $su(3)$ can be obtained simply by reducing the
tensor product of the basic representations $3$ and
$\overline{3}$. In particular it is well known that the tensor
product of the basic representations decompose as follows \cite{georgi}:

\begin{eqnarray}\label{tensor}
  3\otimes 3 &=& 6 \oplus \overline{3}, \cr
  \overline{3}\otimes \overline{3} &=& \overline{6} \oplus 3, \cr
  3\otimes \overline{3} &=& 8 \oplus 1, \cr
 \overline{3}\otimes 3 &=& 8 \oplus 1.
\end{eqnarray}

There is a simple way for decomposing these representations based
on symmetry under permutation. For example the basis states of
$3\otimes 3$, are written as the sum of a symmetric and
anti-symmetric combination, i.e.

\begin{equation}\label{decom33}
  |\mu_i\ra|\mu_j\ra=\frac{1}{2}(|\mu_i\ra|\mu_j\ra +
  |\mu_j\ra|\mu_i\ra) + \frac{1}{2}(|\mu_i\ra|\mu_j\ra -
  |\mu_i\ra|\mu_j\ra)
\end{equation}
The symmetric multiplet forms the basis states of the
representation $6$ and the antisymmetric multiplet that of
$\overline{3}$. In a similar way, one can decompose
$\overline{3}\otimes \overline{3}$ by writing
$|\overline{\mu}_i\ra|\overline{\mu}_j\ra$ as a sum of a
symmetric part ($\overline{6}$) and antisymmetric part ($3$). The
decomposition of $3\otimes \overline{3}$ takes place by
subtracting the trace part from the combination
$|\mu_i\ra|\overline{\mu}_j\ra$ leaving us with an $8$ and a $1$,
i.e.

\begin{equation}\label{decom33}
  |\mu_i\ra|\overline{\mu}_j\ra=(|\mu_i\ra|\overline{\mu}_j\ra -\frac{1}{3}\delta_{ij}
  \sum_{i}|\mu_i\ra|\overline{\mu}_i\ra)+\frac{1}{3}\delta_{ij}
  (\sum_{i}|\mu_i\ra|\overline{\mu}_i\ra).
\end{equation}\\
\\

\end{document}